\documentclass[twocolumn,prl,showpacs]{revtex4}
\usepackage{amsmath}
\usepackage{graphicx}

\newcommand{\eq}{\begin{equation}}
\newcommand{\fine}{\end{equation}}

\begin{document}

\title{Anomalous lack of decoherence of the Macroscopic Quantum Superpositions
based on phase-covariant Quantum Cloning.}
\author{Francesco De Martini$^{1,2}$, Fabio Sciarrino$^{1}$, and Nicol\`{o} Spagnolo$%
^{1}$}

\affiliation{$^{1}$Dipartimento di Fisica dell'Universit\'{a} ''La Sapienza'' and
Consorzio Nazionale Interuniversitario per le Scienze Fisiche della Materia,
Roma, 00185 Italy\\
$^{2}$ Accademia Nazionale dei Lincei, via della Lungara 10, I-00165 Roma,
Italy}

\begin{abstract}
We show that all Macroscopic Quantum Superpositions (MQS) based on
phase-covariant quantum cloning are characterized by an anomalous high
resilence to the de-coherence processes. The analysis supports the results
of recent MQS\ experiments and leads to conceive a useful conjecture
regarding the realization of complex decoherence - free structures for
quantum information, such as the quantum computer.
\end{abstract}

\maketitle

Since the early decades of the last century the counter-intuitive properties
associated with the superposition state of macroscopic objects and the
problem concerning the ''classicality''\ of Macrostates were the object of
an intense debate epitomized in 1935 by the celebrated 
Schr\"{o}dinger's paradox  \cite{Eins35,Schr35}. However, the actual
feasibility of any Macroscopic Quantum Superposition (MQS) adopting photons,
atoms, electrons in SQUIDS \cite{Brun92,Raim01,Leg02,Ourj06} was always
found to be challenged by the very short persistence of its quantum
coherence, i.e. by its overwhelmingly fast ''decoherence''. The latter
property was interpreted as a consequence of the entanglement between the
macroscopic system with the environment \cite{Niel00,Zure03,Dur02}.
Recently, decoherence has received a renewed attention in the framework of
quantum information where it plays a detrimental role\ since it conflicts
with the realization of any device bearing any relevant complexity, e.g. a
quantum computer \cite{Gori07}. In particular, effort has been aimed at
the implementation of \ MQS involving coherent states\ of light, which
exhibit elegant Wigner function representations \cite{Schl01}. Nevertheless
in\ all previous realizations the MQS was found so fragile that even the
loss of a single particle spoils any direct observation of its quantum
properties.

\begin{figure}[t]
\centering
\includegraphics[width=0.45\textwidth]{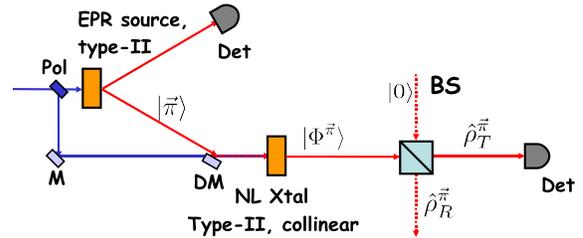}
\caption{Amplified states of the collinear optical parametric
amplifier injected by a single photon qubit generated in a type-II EPR
source. Beam splitter (BS) loss model for the amplified Macro-states
generated by the collinear QI-OPA.}
\label{fig:fig1}
\end{figure}

The present work considers in general a novel type of MQS, one that is based
on the amplification, i.e. or ''quantum cloning'', of a ''Microscopic''\
quantum state (Micro-state) e.g. a single-particle qubit: $|\phi \rangle $ =$%
\;2^{-\frac{1}{2}}\left( |\phi _{1}\rangle +e^{i\varphi }|\phi _{2}\rangle
\right) $. Formally the amplification is provided by a unitary cloning
transformation $\hat{U}$, i.e. a quantum map\ which, applied to the
Micro-state leads to the MQS Macroscopic state (Macro-state): $|\Phi ^{\phi
}\rangle $ $=\hat{U}|\phi \rangle $. In general the amplification can be
provided by a laser - amplifier or by any quantum-injected nonlinear (NL)\
optical parametric amplification (OPA)\ process directly seeded by $|\phi
\rangle $. The \textit{atom-laser} is also a candidate for this process, and
then, within the exciting \textit{matter-wave} context our model can open
far reaching fields of novel scientific and technological endeavour. It
can be shown that $\hat{U}$ can be ''information preserving'', albeit
slightly noisy, and able to transfer in the Macro domain the quantum
superposition character of the single-particle qubit \cite
{DeMa98,DeMa05,Scia05}. Furthermore, unlike most of the other MQS\
schemes, $\hat{U}$, being a fixed intrinsic dynamical property of the
amplifier is not affected in principle by events of scattering of particles
out of the system , i.e. by loss, a process which is generally the dominant
source of decoherence. For the same reasons $\hat{U}$ is largely
insensitive to temperature effects. Let's investigate this interesting
process by the quantum-injected OPA, often referred to as QI-OPA. By this
device, indeed a high-gain phase $(\phi )$-covariant cloning machine seeded
by an entangled EPR photon pair, a Macro-state consisting of a large number
of photons $\mathcal{N}\approx 10^{5}$ was generated \cite
{DeMa08,DeMa08a,Naga07,Scia05,Scia07}. A sketchy draft of the apparatus is
shown in the left part of Figure 1. A polarization $(\vec{\pi})$ entangled
couple of single-photons (A, B)\ is parametrically generated by a standard
Einstein-Podolsky-Rosen-Bohm (EPR) configuration in a NL crystal of BBO\
(beta-barium-borate) cut for Type II phase-mathching and excited by a low
intensity ultraviolet (UV)\ laser beam \cite{DeMa98}. One of the photons,
say A with state $\left| \phi^{\bot} \right\rangle _{A}$, measured by a detector
(Det), provides the trigger signal for the overall experiment. The photon B
with state $\left| \phi \right\rangle _{B}$ nonlocally\ correlated to A, is
injected, via a dichroic mirror (DM) in another BBO\ NL-crystal excited by a
large intensity UV beam. This second NL\ crystal, cut for collinear 3-wave
NL\ interaction provides the large amplification of the injectet seed photon 
$\left| \phi \right\rangle _{B}$ with a NL parametric\ ''gain'' that
attained in the experiment the large value $g\simeq 5$. Precisely its effect
is formally represented by the\ map $|\Phi ^{\phi }\rangle _{B}$ $=\hat{U}%
|\phi \rangle _{B}$ \ with generation of the Macrostate $|\Phi ^{\phi
}\rangle _{B}$, as said. A quantum analysis of the QI-OPA\ interaction will
be given below. As we shall see, the interesting aspect of this process is
that any quantum superposition property of the injected Micro-state $|\phi
\rangle _{B}$ is deterministically mirrored by $\hat{U}$ into the same
property for the Macro-state $|\Phi ^{\phi }\rangle _{B}$. Then the large
fringe ''Visibility'' easily attainable for a Micro-state quantum
interference can be reproduced in the multi-particle MQS regime. In
addition, the property of nonlocal Micro-Micro entanglement $(\left| \phi^{\bot} 
\right\rangle _{A},\left| \phi \right\rangle _{B})\ $correlating the two
EPR\ single-photons is transferred to an entangled Micro-Macro system $%
(\left| \phi^{\bot}  \right\rangle _{A},|\Phi ^{\phi }\rangle _{B})$ then realizing an experimental scheme reproducing exactly the original 1935 Schr\"{o}dinger's proposal
\cite{Schr35}. The experimental demonstration of the\ EPR\ non-separability of a
Micro-Macro entangled ''singlet'' has been reported recently \cite{DeMa08}.
The same procedure can be straightforwardly extended to a Macro-Macro scheme
involving two spacelike separated, entangled Macro-states $(|\Phi ^{\phi
}\rangle _{A},|\Phi ^{\phi }\rangle _{B})$: indeed a novel conceptual
achievement that goes even beyond the Schr\"{o}dinger's paradigm \cite
{DeMa08B}.

The analysis reported in the present letter accounts for the striking
reduction of decoherence of the QI-OPA\ device and is further extended to
consider the apparatus dubbed Orthogonality Filter (OF), which is adopted to enhance the interference Visibility\ implied by the EPR\
entanglement. The decoherence-free QI-OPA\ analyzed by our theory can open
new perspectives for investigating physical systems in a regime close to the
quantum-classical boundary. For instance, it has been recently proposed to
exploit that technique to perform quantum experiments with human-eye
detectors whose ''threshold'' response is simulated closely by the OF \cite
{Brun08,Seka09}.

Let's analyze the resilience to decoherence of this new class of MQS's on
the basis of the concept of \textit{distance} $D(\widehat{\rho },\widehat{%
\sigma })$ in Hilbert spaces between two density operators $\widehat{\rho }\ 
$and $\widehat{\sigma }\;$by further adopting the useful concepts of \ ''%
\textit{pointer states}'', ''\textit{environment induced superselection}'',
''\textit{information flow}''\ \cite{Zure03}. Precisely, In order to
characterize two Macroscopic states $|\phi _{1}\rangle $ and $|\phi
_{2}\rangle $ and the corresponding MQS's: $|\phi ^{\pm }\rangle =\frac{%
\mathcal{N}_{\pm }}{\sqrt{2}}\left( |\phi _{1}\rangle \pm |\phi _{2}\rangle
\right) $, we adopt the following criteria based on the ''Bures metric'': $D(%
\widehat{\rho },\widehat{\sigma })=\sqrt{1-\mathcal{F}(\widehat{\rho },%
\widehat{\sigma })}$ where $\mathcal{F}(\widehat{\rho },\widehat{\sigma })=%
\mathrm{Tr}\left( \sqrt{\widehat{\rho }^{\frac{1}{2}}\widehat{\sigma }%
\widehat{\rho }^{\frac{1}{2}}}\right) \ $ is a ''quantum fidelity'' \cite
{Bure69,Jozs94}. \newline
\textbf{Criterium I)} The ''\textit{Distinguishability}'' between $|\phi
_{1}\rangle $ and $|\phi _{2}\rangle $ can be quantified as $D\left( |\phi
_{1}\rangle ,|\phi _{2}\rangle \right) $. \textbf{Criterium II)} The ''%
\textit{Visibility''}, i.e. ''degree of orthogonality''\ of the MQS's $|\phi
^{\pm }\rangle $ is expressed again by: $D\left( |\phi ^{+}\rangle ,|\phi
^{-}\rangle \right) $. Indeed, the value of the MQS\ visibility depends
exclusively on the relative phase of the component states:$\ |\phi
_{1}\rangle $ and $|\phi _{2}\rangle $. Assume two orthogonal superpositions 
$|\phi ^{\pm }\rangle $: $D\left( |\phi ^{+}\rangle ,|\phi ^{-}\rangle
\right) =1$. In presence of decoherence the relative phase between $\ |\phi
_{1}\rangle $ and $|\phi _{2}\rangle $ progressively randomizes and the
superpositions $|\phi ^{+}\rangle $ and $|\phi ^{-}\rangle $ approach an
identical fully mixed state leading to: $D\left( |\phi ^{+}\rangle ,|\phi
^{-}\rangle \right) =0$. For more details on the criteria, refer to \cite
{DeMa09}. The physical interpretation of $D\left( |\phi ^{+}\rangle ,|\phi
^{-}\rangle \right) $ as ''\textit{Visibility}''\ of a superposition $|\phi
^{\pm }\rangle $ is legitimate insofar as the component states of the
corresponding superposition, $|\phi _{1}\rangle $ and $|\phi _{2}\rangle $
may be defined, at least approximately, as ''\textit{pointer states}''\ or ''%
\textit{einselected states}''\ \cite{Zure03}. These states, within the set
of the eigenstates characterizing any quantum system, are defined as the
ones least affected by the external noise and highly resilient to
decoherence. In other words, the pointer states\ are ''quasi classical''\
states which realize the minimum flow of information from (or to) the System
to (or from) the Environment. They are involved in all criteria of
classicality, such as the ones based on ''purity''\ and ''predictability''\
of the macrostates \cite{Zure03}.

Let's refer to Figure \ref{fig:fig1}. A single-photon state, conditionally
generated with polarization $\vec{\pi}_{\phi }=\frac{1}{\sqrt{2}}\left( \vec{%
\pi}_{H}+e^{\imath \phi }\vec{\pi}_{V}\right) $ by ''spontaneous parametric
down conversion'' in a first NL crystal, is injected into an OPA device. The
interaction Hamiltonian of this process is: $\widehat{\mathcal{H}}%
_{coll}=\imath \hbar \chi \widehat{a}_{H}^{\dag }\widehat{a}_{V}^{\dag }+%
\mathrm{h.c.}$ in the $\left\{ \vec{\pi}_{H},\vec{\pi}_{V}\right\} $
polarization basis, and $\widehat{\mathcal{H}}_{coll}=\frac{\imath \hbar
\chi }{2}e^{-\imath \phi }\left( \widehat{a}_{\phi }^{\dag \,2}-e^{\imath
2\phi }\widehat{a}_{\phi _{\bot }}^{\dag \,2}\right) $ for any
''equatiorial''\ basis $\left\{ \vec{\pi}_{\phi },\vec{\pi}_{\phi \perp
}\right\} $ on the Poincar\'{e} sphere. Two relevant equatorial basis are $%
\left\{ \vec{\pi}_{+},\vec{\pi}_{-}\right\} $ and $\left\{ \vec{\pi}_{R},%
\vec{\pi}_{L}\right\} $ corresponding respectively to $\phi =0$ and $\phi
=\pi /2$. We remind that the \textit{phase-covariant} cloning process
amplifies identically all ''equatorial''\ qubits. The symbols H and V refer
to horizontal and vertical field polarizations, i.e. the extreme ''poles''\
of the Poincar\'{e} sphere. By direct calculation, the amplified states for
an injected qubit $\pi =\left\{ H,V\right\} $ is:$|\Phi ^{\pi }\rangle
=C^{-2}\sum_{i=0}^{\infty }\Gamma ^{i}\sqrt{i+1}\,|(i+1)\pi ,i\pi _{\bot
}\rangle \;$where $C=\cosh g$ , $\Gamma =\tanh g$ and the ket $|n\pi ,m\pi
_{\bot }\rangle $ represents the number state with $n$ photons with $\pi $
polarization and $m$ photons with $\pi _{\bot }$ polarization. The amplified
state for an injected \textit{equatorial }qubit is: 
\begin{equation}
|\Phi ^{\phi }\rangle =\sum_{i,j=0}^{\infty }\gamma _{ij}|(2i+1)\phi
,(2j)\phi _{\bot }\rangle ,
\end{equation}
where $\gamma _{ij}=2^{-(i+j)}C^{-2}\left(
e^{-\imath \varphi }\Gamma \right) ^{i}\left( -e^{\imath \varphi }\Gamma
\right) ^{j}\frac{\sqrt{(2i+1)!}\,\sqrt{(2j)!}}{i!j!}$.

Consider the MQS of the macrostates $|\Phi ^{+}\rangle $, $|\Phi ^{-}\rangle 
$: $|\Psi ^{\pm }\rangle =\frac{\mathcal{N}_{\pm }}{\sqrt{2}}\left( |\Phi
^{+}\rangle \pm i|\Phi ^{-}\rangle \right) $. Due to the linearity of the
OPA process \cite{DeMa05}, it can be easily found: $|\Psi ^{\pm }\rangle
=\left| \Phi ^{R/L}\right\rangle $. Our interest is aimed at the resilience
properties of the QI-OPA\ generated MQS (q - MQS)\ after the propagation
over a lossy channel. This one is modelled by a linear beam-splitter (BS)\
with transmittivity $T$ and reflectivity $R=1-T$ acting on a state $\widehat{%
\rho }$ associated with a single BS\ input mode: Fig.\ref{fig:fig1} \cite
{Loud}. The calculation of the output density matrix $\widehat{\rho }_{T}$
consists of the insertion of the BS\ unitary as a function of $T$ \ followed
by the partial tracing of the emerging field on the loss variables of the
reflected mode.

\begin{figure}[t]
\centering
\includegraphics[scale=.42]{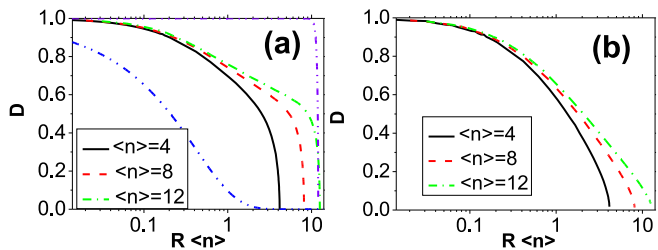}
\caption{(\textbf{a}) Numerical evaluation of the Bures distance $D_q(x)$
between two orthogonal equatorial macro-qubits $|\Phi ^{\protect\phi ,%
\protect\phi _{\bot }}\rangle $ as function of the average lost particle $%
x=R<n>$, plotted in a logarithmic scale. Black continuous line refers to $g=0.8$ and to 
$\langle n\rangle \approx 4$, red dashed line to $g=1.1$ and $\langle n\rangle
\approx 8$, green dash-dotted line to $g=1.3$ and $\langle n\rangle \approx 12$. The
two dash-dot-dot lines correspond to the function $D_{\alpha}(x)$ for coherent state MQS. (\textbf{b}) Numerical
evaluation of the Bures distance between $|\Phi ^{H}\rangle $ and $|\Phi
^{V}\rangle $ for the same values of the gain of (a). }
\label{fig:fig5}
\end{figure}

We have evaluated numerically the \textit{distinguishability} of $\left\{
|\Phi ^{+,-}\rangle \right\} $ through the distance $D(|\Phi ^{+}\rangle
,|\Phi ^{-}\rangle )$ between the states: Fig.\ref{fig:fig5}-(\textbf{a}).
It is found that this property of $\left\{ |\Phi ^{+,-}\rangle \right\} $
coincides with the MQS\ Visibility of $|\Psi ^{\pm }\rangle $, in virtue of
the $\phi $-covariance of the process: $D(|\Psi ^{+}\rangle ,|\Psi
^{-}\rangle )$= $D(|\Phi ^{R}\rangle ,|\Phi ^{L}\rangle )$= $D(|\Phi
^{+}\rangle ,|\Phi ^{-}\rangle )\equiv D_{q}$. The investigation on the
Glauber's states leading to the $\alpha -MQS^{\prime }s$ case \cite{Schl91}:$%
\ |\Phi _{\alpha \pm }\rangle $=$\left( |\alpha \rangle \pm |-\alpha \rangle
\right) [2(1\pm e^{-4|\alpha |^{2}})]^{-%
{\frac12}%
}$ in terms of the ''pointer states'' $|\pm \alpha \rangle \ $leads to the
closed form result:\ $D(|\Phi _{\alpha +}\rangle ,|\Phi _{\alpha -}\rangle )$
=$\sqrt{1-\sqrt{1-e^{-4R|\alpha |^{2}}}}$ $\neq D(|\alpha \rangle ,|-\alpha
\rangle )$=$\sqrt{1-e^{-2(1-R)|\alpha |^{2}}}$. $D(|\alpha \rangle ,|-\alpha
\rangle )$ and $D(|\Phi _{\alpha +}\rangle ,|\Phi _{\alpha -}\rangle )$ are
plotted in Fig.\ref{fig:fig5}-(\textbf{a}) (dash-dot-dot, blue lines)\ as
function of \ the average lost photons: $x\equiv R<n>$. Note that $D(|\alpha
\rangle ,|-\alpha \rangle )$ $\approx 1$ keeps constant, as expected for
pointer states, throughout the values of $x$, up to the maximum value $x=$ $%
<n>$, corresponding to full particle loss, viz to $R\sim 1$. On the contrary
the function $D(|\Phi _{\alpha +}\rangle ,|\Phi _{\alpha -}\rangle )$ drops
from 1 to 0.095 with zero slope upon loss of only one photon: $x=1$, in
agreement with all experimental observations. The \textit{visibility} of the
q-MQS $\left\{ |\Psi ^{+,-}\rangle \right\} $ as well as for the MQS\
components $|\Phi ^{+}\rangle $and $|\Phi ^{-}\rangle $ was evaluated
numerically analyzing the corresponding Bures distance $D_{q}(x)$ as a
function of x. The results for different values of the gain are 
reported in Fig.\ref{fig:fig5}-(\textbf{a}). Note that\ for small values of $%
x$ the decay of $D_{q}(x)$ is far slower than for the coherent $\alpha -MQS$
case. Furthermore, interestingly enough, after a common inflexion point at $%
D_{q}\sim 0.6$ the \textit{slope} of all functions $D_{q}(x)$ corresponding
to different values of $<n>$ increases fast towards the \textit{infinite}
value, for increasing $\ x\rightarrow <n>$ and: $R\rightarrow 1$. All this
means that the q-MQS\ Visibility can be very large even if the average
number $x$ of lost particles is close to the initial total number $<n>$,
i.e. $R\sim 1$. This behavior is opposite to the $\alpha -MQS$ case where
the function $D(R|\alpha |^{2})$ approaches zero with a \ \textit{zero} 
\textit{slope}, as said. Note in Fig.\ref{fig:fig5}-(\textbf{b}) a faster
decay of $D_{q}(|\Phi ^{H}\rangle ,|\Phi ^{V}\rangle )$ due to the fact that 
$|\Phi ^{H}\rangle $ and $|\Phi ^{V}\rangle $ \textit{do not} belong to the
equatorial plane of the Poincar\'{e} sphere, i.e. to the privileged $\phi $%
-covariant Hilbert subspace. This behavior supports all our previous
experimental results, as the recent ones related to the Micro-Macro
nonseparability \cite{DeMa08} and demonstrates the high resilience to
decoherence of the q-MQS\ solution here considered.

The above considerations lead quite naturally to\ an important conjecture.
If the structure of a generic quantum system allows any MQS $|\Theta ^{\pm
}\rangle $=$\frac{\mathcal{N}_{\pm }}{\sqrt{2}}\left( |\Xi ^{+}\rangle \pm
i|\Xi ^{-}\rangle \right) $ and the component interfering Macro-states $|\Xi
^{\pm }\rangle $ to belong to a common covariant Hilbert subspace such as $%
D(|\Theta ^{+}\rangle ,|\Theta ^{-}\rangle )$ = $D(|\Xi ^{+}\rangle ,|\Xi
^{-}\rangle )$, that MQS\ as well as the component Macrostates are
decoherence-free, i.e. bear at least approximately the property of the
''pointer state''. In other words this covariant subspace is the only location
in the world assuring the detectable survival of all MQS.
If the present theorem is true, we could have established a first, useful criterion
for realizing any complex decoherence-free quantum information structure,
e.g. a quantum computer.

\begin{figure}[h]
\centering
\includegraphics[width=0.4\textwidth]{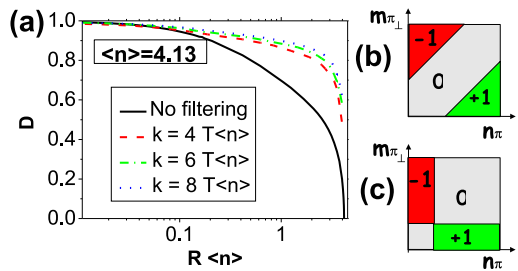}
\caption{(a) Numerical evaluation of the Bures distance between two
orthogonal equatorial O-filtered macro-qubits for different values of the
threshold $k$ ($g=0.8$). (b) Bidimensional Fock-space represation of the
action of the O-Filter device. The various outcomes are assigned depending
on the difference between the measured photon number in the two polarization 
$n_{\protect\pi}$ and $m_{\protect\pi_{\bot}}$. The threshold $k$ allows to
vary the dimension of the selected zone. (c) Human-eye detection scheme
presented in \protect\cite{Seka09}.  }
\label{fig:fig3}
\end{figure}

The distinguishibility between the pairs of states $\left\{ |\Phi ^{\phi
}\rangle ,|\Phi ^{\phi _{\bot }}\rangle \right\} $ after losses was greatly
improved by the use of the (OF) device , allowing the demonstration of the
Micro-Macro entanglement \cite{DeMa08}. The \textit{local} POVM like (OF)
technique \cite{Pere95} selects the events for which the difference between
the photon numbers associated with two orthogonal polarizations $|m-n|>k$,
i.e. larger than an adjustable threshold, $k$ \cite{Naga07}. By this method
a sharper discrimination between the output states $|\Phi ^{\phi }\rangle $
e $|\Phi ^{\phi _{\bot }}\rangle $ is achieved. This technique presents close analogies with the human-eye
detection scheme presented in \cite{Seka09,Brun08}. Indeed, as shown
in Fig.\ref{fig:fig3}-(b-c), the two schemes select similar zone of the
bidimensional Fock-Space proper of the QIOPA output states. We analyzed the discriminating effects of the OF device on the
orthogonal macroqubits after losses $\hat{\rho}_{T}^{+}$ and $\hat{\rho}%
_{T}^{-}$. In Fig.\ref{fig:fig3}-(a) the results
of the numerical analysis carried out for $g=0.8$ and different values of $k$%
\ are reported. Note the increase of the value of $D(x)$, i.e. of the
q-MQS\ Visibility, by increasing $k$. Of course all this is achieved at the
cost of a lower success probability. All these results are in
agreement with a recent MQS experiment involving a number of photons in
excess of $5\times 10^{4}$ , indeed the only experiment so far exploiting
this technique \cite{DeMa08}. There a q-MQS Visibility $>50\%\ $was measured
upon a loss of more than $R\sim 90\%$ of the QIOPA\ generated particles, at 
\textit{room temperature}.

The efficiency of the transfer of classical or quantum information in the
interactive dynamics involving the paradigmatic quantum - statistical
combination (\textit{System +Environment}) is at the focus of the present
investigation. In order to gain insight into the general picture and to
support the congruence of our final conclusions we find useful to relate
here the various aspects of the cloning process with the current MQS\
physical models\cite{Zure03}.

1) The ''\textit{System}'' in our entangled OPA scheme is represented by the
assembly of $N+1$ photon particles associated with the entangled ''singlet''
macrostate $2^{-%
{\frac12}%
}(|\phi _{\perp }\rangle |\Phi ^{\phi }\rangle -|\phi \rangle |\Phi _{\perp
}^{\phi }\rangle )$ generated by the combined EPR - quantum cloning
apparatus.

2) The flow of (classical) ''noise information'' directed from the ''\textit{%
Environment}'' towards the \textit{System} is generally provided by the
unavoidable squeezed-vacuum noise affecting the building up of the
macrostate $|\Phi ^{\phi }\rangle $ within the OPA\ process. \ As already
stressed, the ''\textit{optimality}'' of the $\phi $ - covariant cloning
generally implies, and literarly means, that the flow of classical noise is
the \textit{minimum} allowed by the principles of quantum mechanics, i.e. by
the ''\textit{no-cloning theorem}'' \cite{Scia05,Scia07}.

3) The flow of quantum information directed from the System towards the
Environment is provided by the controlled \textquotedblright decoherence in
action\textquotedblright\ provided by the artificial BS-scattering process
and by the losses taking place in all photo-detection processes. We have
seen that by the use of the OF, or even in the absence of it, the
interference phase-distrupting effects caused by the adopted artificial
decoherence can be efficiently tamed and even cancelled for the
\textquotedblright equatorial\textquotedblright\ macrostates and for their
quantum superpositions.

4) The selected $\phi $ - covariant cloning method considered here allows to
define the ''equatorial'' plane as a privileged ''\textit{minimum noise -
minimum decoherence}''\ Hilbert subspace of the quantum macrostates that,
according to our decoherence model, exhibit simultaneously the maximum
allowed \textit{Distinguishability} and \textit{Visibility}.

5) In any cloning apparatus a unitary $\hat{U}$ connects all physical
properties belonging to the micro-world to the corresponding ones belonging
to the macrosopic ''classical''\ world. Any lack of perceiving this close
correspondence, for instance in connection with the realization of the 1935 Schr\"{o}dinger's proposal must be only attributable to the intrinsic
limitations of our perceiving senses, of our observational methods or of our
measurement apparata. In other words, at least in our case, the two worlds
are deterministically mirrored one into the other by the map $\hat{U}$ \
which is provided by quantum mechanics itself. This is the key to understand
our results.

6) The q-MQS based on the cloning process is not a ''thermodynamic'' system
as its dynamics and decoherence do not depend on the temperature T. It
rather belongs to, and indeed establishes a first and most insightful
physical model of, the class of the ''\textit{parametrically - driven, open
quantum - statistical systems}'' that have been recently invoked\ to provide
and sustain\ extended long - range nonlocal coherence processes in complex
biological photosyntetic systems \cite{Cai08}. \ We acknowledge 
useful discussions with Chiara Vitelli.

\end{document}